\documentclass[prl,onecolumn,preprint,amsmath,amssymb,floatfix,superscriptaddress]{revtex4}
\usepackage[dvipdfmx]{graphicx,epsfig}
\usepackage{rotate}
\usepackage[dvipdfmx]{color}
\usepackage{bm}
\usepackage{amsmath,amssymb}
\usepackage{url}

\begin{document}

\title{Contribution of hidden modes to nonlinear epidemic dynamics in urban human proximity networks}

\author{Naoya Fujiwara}
\email{fujiwara@csis.u-tokyo.ac.jp}
\affiliation{Center for Spatial Information Science, The University of Tokyo, 277-8568 Chiba, Japan}
\affiliation{Institute of Industrial Science, The University of Tokyo, 153-8505 Tokyo, Japan}
\author{Abhijeet R. Sonawane}
\affiliation{Channing Division of Network Medicine, Brigham and Women's Hospital, Boston MA 02115, USA}
\affiliation{Harvard Medical School, Boston MA 02115, USA}
\affiliation{Institute of Industrial Science, The University of Tokyo, 153-8505 Tokyo, Japan}
\author{Koji Iwayama}
\affiliation{Faculty of Agriculture, Ryukoku University, Shiga 520-2194, Japan}
\affiliation{Institute of Industrial Science, The University of Tokyo, 153-8505 Tokyo, Japan}
\author{Kazuyuki Aihara}
\affiliation{Institute of Industrial Science, The University of Tokyo, 153-8505 Tokyo, Japan}

\begin{abstract}
Recently developed techniques to acquire high-quality human mobility data allow large-scale simulations of the spread of infectious diseases with high spatial and temporal resolution.
Analysis of such data has revealed the oversimplification of existing theoretical frameworks to infer the final epidemic size or influential nodes from the network topology.
Here we propose a spectral decomposition-based framework for the quantitative analysis of epidemic processes on realistic networks of human proximity derived from urban mobility data.
Common wisdom suggests that modes with larger eigenvalues contribute more to the epidemic dynamics.
However, we show that hidden dominant structures, namely modes with smaller eigenvalues but a greater contribution to the epidemic dynamics, 
exist in the proximity network.
This framework provides a basic understanding of the relationship between urban human motion and epidemic dynamics, and will contribute to strategic mitigation policy decisions.

\end{abstract}

\maketitle

Epidemics of infectious diseases in the human population, e.g. 
the SARS outbreak in 2002--2003 \cite{pccc03}, 2009 H1N1 influenza pandemic \cite{fdch09}, and Ebola outbreak  \cite{ggas14}, can be a serious factor in human mortality, and have a significant socio-economic impact in terms of  reducing a population's healthy years because of morbidity.
In recent years, the mathematical modelling of the outbreak and spread of infectious diseases has mainly been performed using two approaches: agent-based models
\cite{egkm04,siyt13,fccf05,am09,ma10,fcfc06,hfel08} and  structured metapopulation models \cite{wcsa11,cpv07,bv11,ptc12,cbbv06,cbbv07,bcgh09,sd95,v07,wmmd05,reg99,rl85}. Both techniques incorporate real or synthetic data on the long-range mobility and migration of populations,
but vary in their granularity of the population,
 from individuals to  sub-groups of society. 
Various spatial scales have been studied, ranging from specific geographic locales such as cities \cite{egkm04,siyt13} to nations \cite{fccf05,am09,ma10,r07,geg03} and intra-/inter-continental regions \cite{cbbv06,cbbv07}. 

In particular, it is imperative to study the mechanisms of an epidemic's spread in urban scenarios,
considering restrictions on available space  
and transportation methods (e.g. traffic fluxes, travel routes, timescales of modes of transport), which result in high density and heterogeneous contact patterns among residents. These factors make the populace vulnerable to epidemics of diseases that spread through exhaled aerosols, e.g. \cite{t09}.
To understand the details of epidemic dynamics within cities, it is crucial to exploit the heterogeneous contact patterns derived from human mobility data.
Fortunately, recent technical advances in data acquisition methods enable us to
incorporate human mobility data obtained from various sources \cite{oshs07,cvbc10,ssku11} into  mathematical models \cite{cbbv07}.

Network epidemiology \cite{mh13,wmmd05,dfhj11,pcvv15,kghl10,bh13}  is a useful means of
uncovering the dependence of the contagion dynamics 
 on the heterogeneous contact pattern. 
The  approach of tracking connections between people provides a comprehensive
 framework with which to study the effects of the underlying network of people in the region of interest (e.g. cities).
  The present understanding of epidemic or social contagion processes has been bolstered by many network-based studies \cite{cbbv07,skll10,mh13}, which treat cities as social petri-dishes \cite{kl10}.
 
Analytical approaches such as heterogeneous mean field (HMF) theory \cite{pv01,mpv02,bpv03,ywrbs07}  and moment closure approximation \cite{k99,ek02,bs02} provide an important bridge between the network topology and the spread of epidemics in complex networks.
For example, the epidemic threshold is absent in scale-free networks with no degree correlation \cite{pv01}.
However, certain simplifications made to obtain analytical results, e.g.  neglecting the degree correlation  in the HMF \cite{mpv02}, could be superfluous,  and these approximations may not necessarily hold in realistic networks. 
 Therefore, a novel analytical approach is required to better understand the spread of epidemics in heterogeneous populations.

\section*{Proximity networks and the SIR model }
In this paper, we outline a framework for modelling a proximity network from transportation data, and propose an  analytical method to estimate the infection probability of each individual from stochastic simulations. The prevalence, i.e. the average number of recovered agents who were once infected before the spread of the epidemic subsides, is of particular interest.

We can construct an individual-based network using a dataset of human mobility. The \textit{People flow data} 
 \cite{ssku11} record the one-day movement of individuals living in the Kanto region of Japan, which includes the Tokyo metropolitan area.
  Details of the dataset are given in the Methods section. 
We randomly chose the mobility track of $N$ individuals from this dataset.
 Time-dependent  proximity networks were then constructed by connecting individuals when they came within a certain geographical proximity, defined by a distance threshold $D$ (Fig.~\ref{fig:map}).
 Let us denote the adjacency matrix of the proximity network as $A(t)$.
All results in the main text were obtained with $D=1,000$~m and $N=10,000$. Note that we observed the percolation-like transition and the scaling relation of the giant cluster component size ratio as $S(N,D)/N \sim f(N^{0.605}D)$.  Using this scaling relation, $D=1,000$~m with $N=10,000$ corresponds to $D\approx 8.3$~m with the real population in the corresponding region, where $N\approx 27,500,000$. A detailed discussion can be found in Supplementary Information SI1.

We simulated the disease propagation using the agent-based SIR model \cite{ys11} on the above-mentioned network, which assigns 
 susceptible (S), infected (I), or recovered (R) states to each agent.
We assumed that each agent repeats the same trip pattern every day in the same way as recorded in the People flow data.
When a susceptible agent is connected with an infected one, 
the former is infected with probability $\beta \Delta t$ for a time interval $\Delta t$.
If a susceptible agent is connected with  $d$ infected agents, they are infected with probability $1 - (1-\beta \Delta t)^d$ \cite{mh13}.
Further details of the stochastic simulation are given in the Methods section.
In this study,  
$\beta \Delta t$ is sufficiently small which enables us to perform the simulation with the discretised time step.
An infected agent recovers with probability $\mu \Delta t$ over this time interval.
Once $A(t)$, $\beta$, and $\mu$ are given, stochastic simulations can be conducted to determine the epidemic dynamics.
We denote the probabilities that agent $j$ is in state S, I, or R at time $t$ as $s_j(t)$,
$i_j(t)$, and $r_j(t)$, respectively; 
the equality $s_j(t) + i_j(t) + r_j(t) =1$ holds.
We define the infection probability of agent $j$ as $r_j(\infty)$, because a recovered agent has experienced the infected state before recovering.

The epidemic dynamics can be characterized by three different stages  \cite{bbv08}: the initial stage, in which
stochastic fluctuations are dominant, the exponential growth stage, and the final stage, where nonlinearity suppresses  the further spread of the disease.
The latter two stages representing the dynamics of the outbreak are of particular interest.
In these stages, the time evolution of the epidemic can be approximated by the following deterministic differential equations:
\begin{align}
\begin{split}
\frac{d s_j}{dt} &= - \beta s_j \sum_{k=1}^N \overline A_{jk} i_k,\\
\frac{d i_j}{dt} &= \beta s_j \sum_{k=1}^N \overline A_{jk} i_k -\mu i_j, \\
\frac{d  r_j}{dt} &= \mu i_j.
\label{eq:mf}
\end{split}
\end{align}
Here, we approximate the time-dependent adjacency matrix $A(t)$ with its time averaged form  $\overline{A}$ \cite{svbc11}.  This averaging gives a sufficient approximation of the time evolution of $s_j$, $i_j$, and $r_j$ if the parameters $\beta$ and $\mu$ are sufficiently small, as discussed in  Supplementary Information SI2. We also assume that node $j$ being susceptible and node $k$ being infected are statistically independent events \cite{n10}. Although this assumption does not hold for small networks \cite{ys11}, we have verified that the numerical solution of equation~(\ref{eq:mf}) gives a good approximation of the dynamics if $D$ is much greater than the percolation transition point and the giant cluster component is sufficiently large.

One of the most common methods used to analyse the  contagion processes within networks is   HMF \cite{pv01,mpv02,bpv03,ywrbs07}.
This approach assumes that, in a statistical sense, nodes  grouped by the degrees behave in the same way.
Under this assumption, one can derive the equation for the infection probability of each agent,
and, in the absence of degree correlation, the epidemic threshold, i.e. the critical value of the epidemic's spread, can be represented as a function of  the first and second moments of the degree distribution \cite{mpv02}. 
One question that persists is whether such an approximation is valid in realistic networks of human contact.
Actually, the accuracy of HMF depends on factors such as the mean degree and first-neighbour degree \cite{gmwp12}.
For the simplest case of the absence of degree correlation, where HMF is often applied, the component of the first eigenvector is proportional to the degree \cite{bbv08}.
However, this assumption does not hold for the proximity network (Fig.~\ref{fig:series}a).
This highlights the need to develop an analytical framework 
to identify  people with a high infection probability along with the overall infection propagation pattern.
In this study, we have developed a method based on spectral decomposition and mode truncation which demonstrates that a small number of dominant modes, which may not necessarily have the largest eigenvalues, give a higher contribution in the prevalence of epidemics.

\section*{Spectral analysis}

\begin{figure*}[htbp]
\vspace{-1cm}
\centering
\begin{minipage}{0.95\hsize}
\includegraphics[width= 0.95 \textwidth]{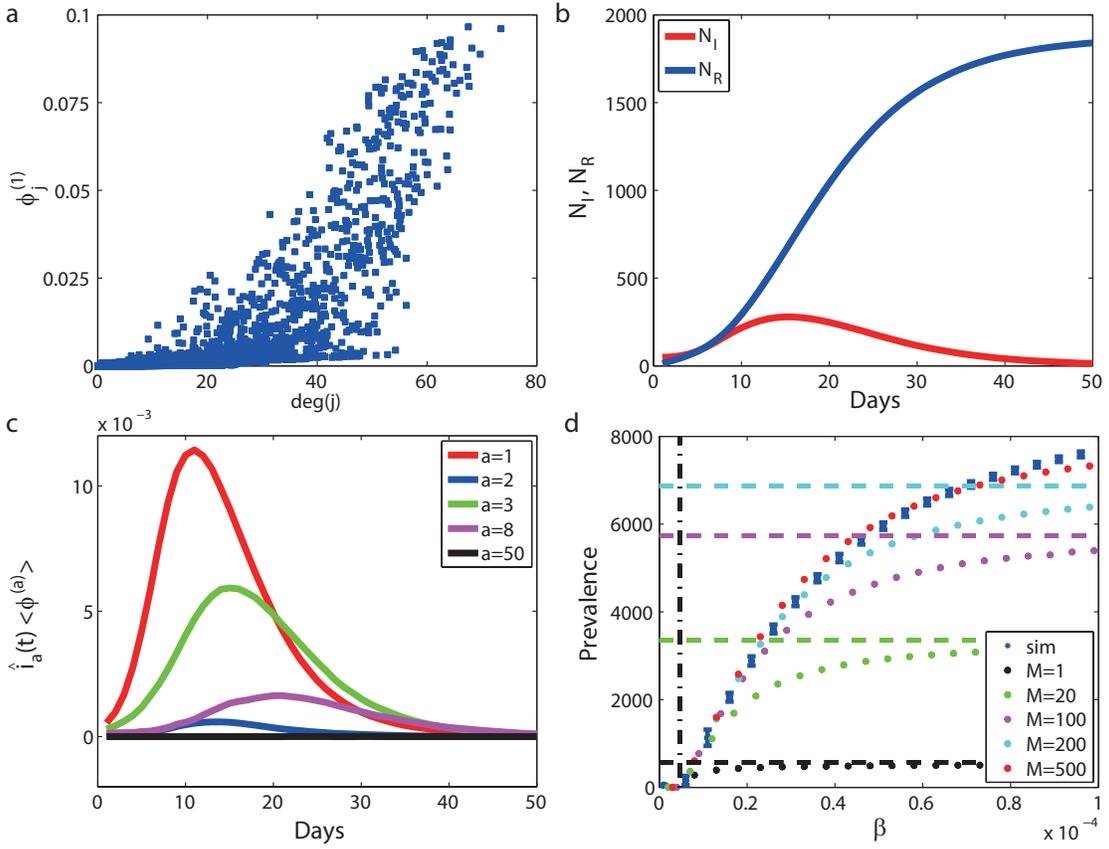}
\end{minipage}
\caption{\label{fig:series}
\label{fig2}
\textbf{Spectral method for the analysis of epidemic spreading.}
a. The first eigenvector component versus the degree of $\overline A$.
For networks with no degree correlation, the component of the first eigenvector 
is proportional to the degree, but this result shows that this is not the case for the proximity network.
b. Time evolution of the average numbers of  infected $N_I$ and recovered $N_R$ people in the stochastic simulation.
The number of infected agents increases exponentially, and then decays because of the nonlinear effect.
c. Time evolution of $\hat i_a(t) \langle \phi^{(a)} \rangle$. Initially, all coefficients grow exponentially, and the mode with the largest eigenvalue $\hat i_1(t) \langle \phi^{(1)} \rangle$ grows fastest.
In the later stages of infection, other modes $\hat i_a(t)$ ($a\neq 1$) grow and $\hat i_1(t)$ decays faster than the others. Hence, the other modes are not negligible. 
However, modes with small eigenvalues are negligible because their amplitude is sufficiently small.
Therefore, much smaller than $N$ (but not one) modes should be taken into account.
d. Dependence of the prevalence on the infection rate $\beta$. 
Results of the stochastic simulation (blue) and numerical solutions of the mode truncated equation (\ref{eq:self-conf_eig}) with different numbers of modes, $M$, are shown.
The horizontal dashed lines represent the cumulative contributions for the all-infected case $\Gamma_M^{\rm all}$.
The vertical broken line at $\beta_c = \mu / \lambda_1 \approx 4.6 \times 10^{-6} \ \rm{min}^{-1}$ is the transition point estimated from the largest eigenvalue of the averaged adjacency matrix $\overline A$.
}
\end{figure*}

In this section, we present a method for analysing the final epidemic size based on the spectral analysis of the averaged matrix $\overline{A}$, and elucidate the relevant dynamics of the epidemic's spread.
The node-wise probabilities $s_j(t)$, $i_j(t)$, and $r_j(t)$ can be expanded as
\begin{align}
\begin{split}
s_j(t) = \sum_{a=1}^N \hat s_a(t) \phi^{(a)}_j,\ 
i_j (t) = \sum_{a=1}^N \hat i_a(t) \phi^{(a)}_j,\ 
r_j (t) = \sum_{a=1}^N \hat r_a (t) \phi^{(a)}_j,
\end{split}
\label{eq:expansion}
\end{align}
where $\phi^{(a)} _j$ denotes the $a$th eigenvector of $\overline{A}$ associated with the 
eigenvalue $\lambda_a $. 
All eigenvalues of a real symmetric matrix $\overline{A}$ are real, and can be labeled in descending order as 
$\lambda_1 \ge \lambda_2 \ge \ldots \ge \lambda_N$. 
The normalization condition $\sum_{j=1}^N (\phi^{(a)}_j)^2  = 1$ is adopted.
Since $\overline A$ is symmetric, the left and right eigenvectors coincide, and the expansion coefficients can be obtained with $\hat s_a(t) = \sum_{j=1}^N s_j(t) \phi_j^{(a)}$, $\hat i_a(t) = \sum_{j=1}^N i_j(t) \phi_j^{(a)}$, and $\hat r_a(t) = \sum_{j=1}^N r_j(t) \phi_j^{(a)}$, respectively.
Equation (\ref{eq:mf}) can be rewritten in terms of these expansion coefficients (Supplementary Information SI3).
The time evolution of the average number of infected and recovered agents, $N_I (t) \equiv \sum_{j=1}^N i_j(t)$ and $N_R (t) \equiv \sum_{j=1}^N r_j(t)$, is plotted in Fig.~\ref{fig:series}b. The time series of $i_j(t)$ is converted to that of $\hat i_a(t) $ as in Fig.~\ref{fig:series}c.

In the exponential stage of the epidemic spreading, 
 $1-s_j$,
 $i_j$, and $r_j$ are small.
The exponential growth 
of each coefficient $\hat i_a(t)$ can be described by the following linearised equation \cite{wcwf03}
\begin{align}
\frac{d \hat i_a}{dt} &\approx ( \beta \lambda_a  -\mu ) \hat i_a
\label{eq:lin_i}
\end{align}
(see Supplementary information SI4 for the derivation of this expression from the nonlinear mode coupling equation).
Hence, $\hat i_a(t) = \hat i_a(0) e^{(\beta \lambda_a - \mu)t}$ holds.
It is evident from this equation that a mode with a larger eigenvalue $\lambda_a$ grows faster in this stage.
In particular, 
the exponential growth rate of the number of infected agents is characterised by $\beta \lambda_1-\mu$.
The stochastic simulation of the agent-based model verifies that $\hat i_1$ is indeed dominant in the initial stage (Fig.~\ref{fig:series}c). 
In addition, the epidemic threshold is related to $\lambda_1$ as $\beta_c = \mu / \lambda_1$ (Fig.~\ref{fig:series}d) \cite{wcwf03}.

\begin{figure*}[htbp]
\vspace{-1cm}
\begin{center}
\begin{minipage}{0.95\hsize}
\includegraphics[width=\textwidth]{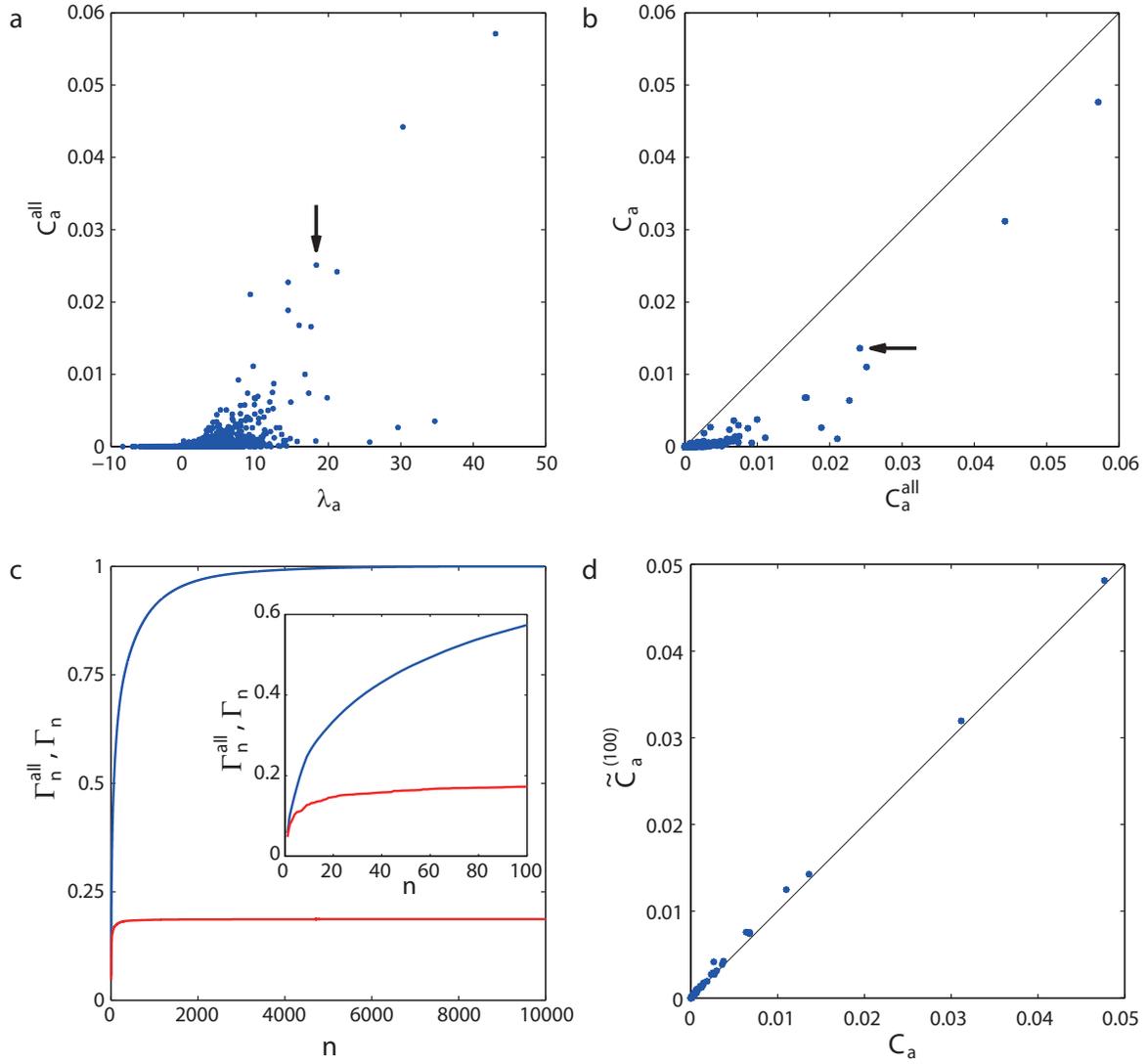}
\end{minipage}
\caption{\label{fig:final_size}
\textbf{Contribution of modes and verification of the final size equation.}
a. Contribution for the all-infected case $C_a^{\rm all}$ versus eigenvalue $\lambda_a$. 
Although positively correlated, $C_a^{\rm all}$ is not a monotonically increasing function of $\lambda_a$.
Arrows in this panel and panel b indicate the mode $a=8$, which has smaller eigenvalue but larger contribution. The time evolution of this mode is depicted in Fig.~\ref{fig:series}c.
b. Contribution $C_a$ versus $C_a^{\rm all}$ obtained from the stochastic simulation. 
They are positively correlated, hence modes giving a high contribution for the all-infected case are important for the case of smaller prevalence.
c. Cumulative contributions $\Gamma_n^{\rm all}$ (blue) and $\Gamma_n$ (red) versus the number of modes $n$. An enlargement is shown in the inset. Contributions are heterogeneous, i.e. 10\% of the total modes contribute approximately 90\% in the all-infected case.
d. Comparison of $C_a$ obtained by the stochastic simulation and that obtained by the numerical solution of equation~(\ref{eq:self-conf_eig}) $\tilde C_a^{(100)}$ computed using $100$ modes.
}
\end{center}
\end{figure*}

However, as shown in Fig.~\ref{fig:series}c, the coefficients $\hat i_a(t) \langle \phi^{(a)} \rangle$ ($a=2,3,\ldots, N$) are comparable to $\hat i_1(t) \langle \phi^{(1)} \rangle$  in the later stages of the epidemic, where $\langle v \rangle = \sum_{j=1}^N v_j /N$ is the node-wise average of a vector $v$.
The importance of each mode needs to be assessed to ascertain its dominance in the spreading mechanism. This can be achieved by introducing a quantifier of the  \textit{contribution} of each mode to the prevalence as
\begin{align}
C_a \equiv \hat r_a(\infty) \langle \phi^{(a)} \rangle.
\end{align}
The definition of this contribution enables us to account for the prevalence as the sum of the contributions of all modes,
 $\sum_{j=1}^N r_j (\infty) = \sum_{a=1}^N C_a$.
It is worth introducing the contribution $C_a^{\rm all}$ for the case where all agents are infected, i.e. $r_j(\infty) = 1$ for all $j$. Since the coefficients satisfy $\hat r_a (\infty) = N\langle \phi^{(a)} \rangle$ in this case, one can easily verify that $C_a^{\rm all} = N \langle \phi^{(a)} \rangle^2$ holds.
Interestingly,  the contribution $C_a^{\rm all}$ does not monotonically increase with the eigenvalue, although a positive correlation can be observed (Fig.~\ref{fig:final_size}a). 
 Since $C_a$ and $C_a^{\rm all}$ are  functions of the average of the eigenvector components, a mode associated with a smaller eigenvalue may contribute more than one with a larger eigenvalue. 
Although the larger eigenvalue denotes faster growth in the exponential regime, it does not necessarily mean a larger contribution to the prevalence.

It is interesting to note that the ``hidden'' important structures, corresponding to modes 
that make a large contribution to the prevalence but have smaller eigenvalues, are unravelled through the spectral analysis. It is still possible to detect such modes when the prevalence is much smaller than the all-infected case (Fig.~\ref{fig:final_size}b). 
The heterogeneity in the contribution is evident from the cumulative contribution $\Gamma_n \equiv \sum_{a=1}^{n} C_{\sigma(a)}$ and $\Gamma_n^{\rm all} \equiv \sum_{a=1}^{n} C_{\sigma(a)}^{\rm all}$  (Fig.~\ref{fig:final_size}c). 
Here, the modes are sorted such that $C_{\sigma(1)}^{\rm all} \ge C_{\sigma(2)}^{\rm all} \ge \cdots \ge C_{\sigma(N)}^{\rm all}$, where $\sigma(a)$ denotes the permutation of the index set $\{ 1,2,3,\ldots ,N\}$.
Note that $\Gamma_N^{\rm all} = 1$ holds, because the infection pattern is represented using all modes.
Figure \ref{fig:final_size}c suggests $\Gamma_{1000}^{\rm all} \approx 0.9$ for the all-infected case, indicating that approximately 90\% of the prevalence is described by the top 10\% of modes for the present proximity network. This figure indicates that the number of modes needed to describe the prevalence decreases if the prevalence is smaller than $1$.

The above observations led us to the idea of describing the epidemic dynamics
with a small number of dominant modes that give a high contribution.
In the case of the agent-based SIR dynamics, we can derive the truncated equation for the contributions at the final time with $M$ ($ \le N$) modes, $\tilde C_{\sigma(a)}^{(M)}(\infty)$,  as 
 \begin{multline}
\tilde C_{\sigma(a)}^{(M)}(\infty) 
=
  N\langle  \phi^{(\sigma(a))} \rangle ^2 -  \sum_{j=1}^N s_j(0) \phi_j^{(\sigma(a))} \langle \phi^{(\sigma(a))} \rangle\\
  \times \exp \left\{ -\frac{\beta}{\mu}  \sum_{b=1}^M\lambda_{\sigma(b)} [\tilde C_{\sigma(b)}^{(M)} (\infty) - \tilde C_{\sigma(b)}^{(M)}(0)] 
  \frac{\phi_j^{(\sigma(b))} }{ \langle \phi^{\sigma (b)} \rangle} \right\},
    \label{eq:self-conf_eig} 
\end{multline}
where $a = 1,2,\ldots,M$.
This equation is derived  in the Methods section. This transcendental equation approximates the epidemic dynamics with fewer effective degrees of freedom. If we use all the modes, i.e. $M=N$, 
the solution of equation (\ref{eq:self-conf_eig}) coincides with 
the solutions of equation~(\ref{eq:mf}) for $t \rightarrow \infty$. 
Furthermore, the solution of this equation  
computed with Newton's method provides a good description of the contribution obtained by the stochastic simulation (Fig.~\ref{fig:final_size}d).
We further note that considering fewer than $N$ modes suffices to approximate the prevalence 
as obtained by the stochastic simulation (Figs.~\ref{fig:series}d,~\ref{fig:final_size}c).
Hence, this procedure replaces the need to solve an ordinary differential equation in $2N$ independent variables (\ref{eq:mf}) 
with  the requirement to use information from the $M$ dominant modes.

As discussed above, the essence of  theoretical analyses of epidemic spreading within networks is to infer the dynamical information from the network topology. 
Conventionally, HMF \cite{pv01,mpv02,bpv03,ywrbs07} uses  degree distributions
and degree correlations 
as topological information.
In contrast, our approach uses  spectral information, which can take  more general properties of the network topology into account.
 As described in the Methods, the final size equation in HMF \cite{mpv02} is derived from equation~(\ref{eq:self-conf_eig}) under the following simple assumptions: (i) degree correlation is absent,  and (ii)  the first mode almost perfectly governs the dynamics.
As shown in Fig.~\ref{fig:series}a, the first assumption is not satisfied in the proximity network.
Moreover, it is evident from Figs.~\ref{fig:series}d and \ref{fig:final_size}a that considering only the first mode is not sufficient to account for the whole epidemic dynamics, as $C_1^{\rm all}$ is less than 6\%.
The contribution of other dominant modes must be taken into account to quantify the epidemic dynamics (Fig.~\ref{fig:final_size}d).
Thus,  HMF with no degree correlation is not a sufficient approximation for the proximity network.
It would be possible to improve the approximation by considering the degree correlation in HMF \cite{bpv03}, but it is generally difficult to include information about the second-nearest neighbours.
In contrast, the proposed approach provides the flexibility to 
improve the approximation by 
using an arbitrary number of dominant modes.

\section*{Identification of the spatio-temporal hot spots of infection}
\begin{figure*}[hbtp]
\begin{center}
\begin{minipage}{0.95\hsize}
\includegraphics[width=\textwidth]{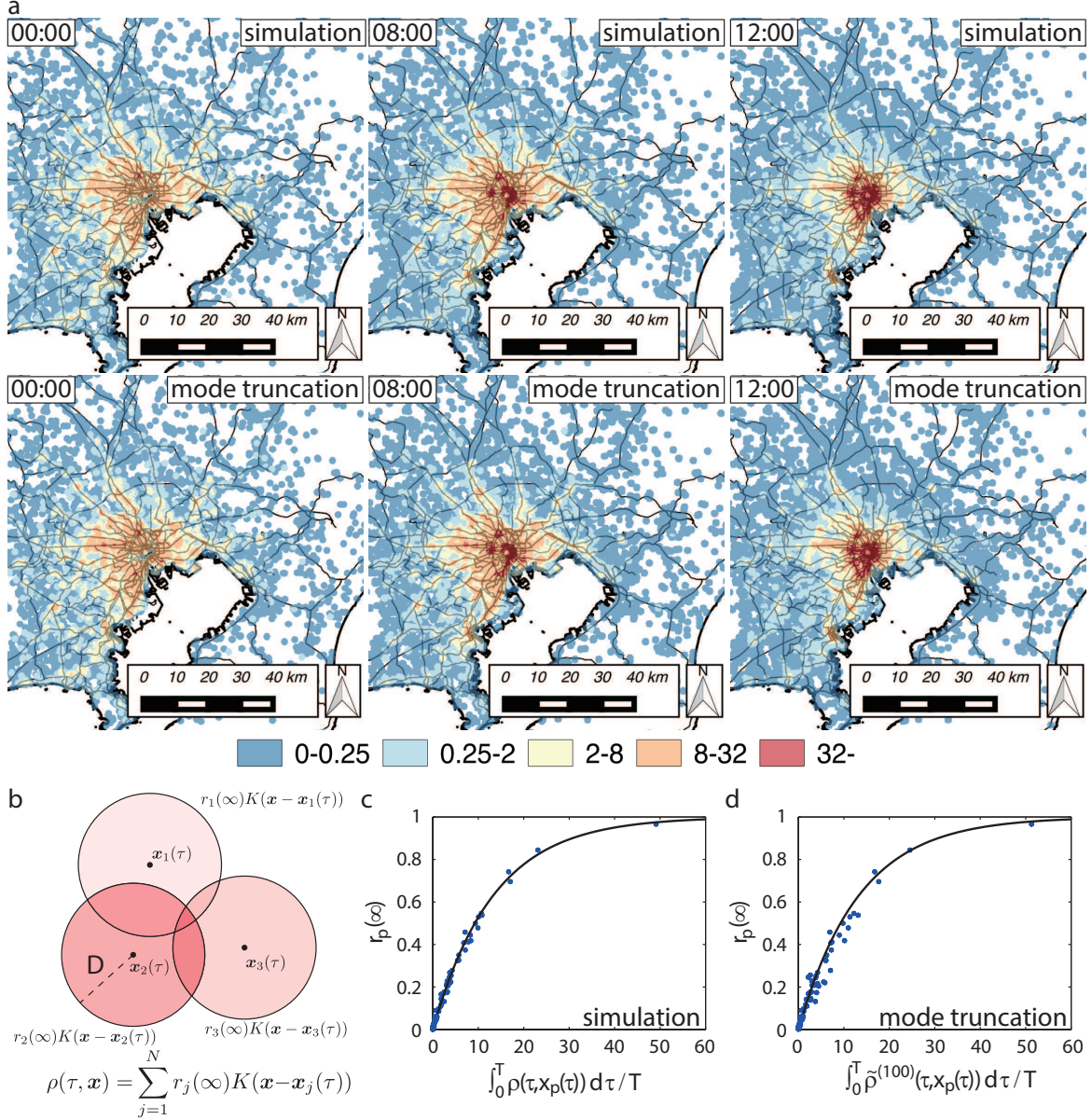}
\end{minipage}
\caption{\label{fig:risk}
\textbf{Identification of spatio-temporal hot spots and estimation of infection probability of test agents.}
a. Spatial distribution of the risk factor $\rho(\tau,\bm x)$ of the proximity network at 00:00, 08:00, and 12:00, respectively. The upper and lower rows give the results of the stochastic simulation and those of the mode truncated equation with $100$ modes, respectively.
b. Schematic diagram to compute the risk factor $\rho(\tau,\bm x)$.
c. Infection probability of the $p$th test agent $r_p(\infty)$ versus integral of the risk factor along the path of the agent $\bm x_p(\tau)$. Here, $T=1,440 $~min is the normalisation constant.
Dots are the numerical results, and the line is the theoretical prediction 
$1- \exp\left[-  \frac{\beta}{\mu T}  \displaystyle \int_0^T \rho(\tau,\bm x_p(\tau) ) d\tau \right]$.
Details of the derivation of this equation is given in Supplementary Information SI5.
d. The same as panel c, but the horizontal axis is the time integral of $\tilde \rho^{(100)}$ obtained from the mode truncation with $100$ modes.
}
\end{center}
\end{figure*}

Another feature of epidemic spreading in proximity networks is the strong heterogeneity of the infection risk in space and time.  It is  important to identify such locations that have a high probability of infection, namely ``hot spots'',  for the mitigation and management of the epidemic. The spatial distribution of the hot spots can be mapped to provide an insight into the nature of epidemic spreading in urban areas (Fig.~\ref{fig:risk}). More importantly, to uncover the relevant epidemic dynamics, such identification should be conducted with fewer parameters than the number of agents. 

For the present model to be predictive, let us infer the  infection probability of a test agent from the mobility pattern of other agents. 
The probability of infection of this agent depends on the presence of other infectious agents in the vicinity, which in turn varies from region to region and at different times of day,  $0 \le \tau \le T = 1,440 \ \rm{min}$. Let $\bm x = (x,y)^T$ be a position vector and $\bm x_j (t)$  be the position of  agent $j$ at time $t$. We introduce a kernel function $K(\bm x)$ signifying the presence or absence of any interaction between two agents based on the distance between them. Since agents are connected within the threshold $D$, the Heaviside step function $K(\bm x) = \Theta(|\bm x| -D)$ is adopted as the kernel function. This allows us to define a spatio-temporal risk factor of infection $\rho(\tau,\bm x)$  as
\begin{align}
	\rho(\tau,\bm x) \equiv \sum_{j=1}^{N}r_j(\infty)K(\bm x - \bm x_j(\tau)),
\end{align}
 which 
can be interpreted as the weighted sum of probabilities $r_j(\infty)$. 
Integrating $\rho(\tau,\bm x_p(\tau))$ along the trajectory of the test agent $\bm x_p(\tau)$, one can obtain the infection probability of this agent. 
Details of the derivation are discussed in Supplementary Information SI5. 
Figure \ref{fig:risk}a depicts the spatial distribution of the risk factor $\rho(\tau,\bm x)$ across the  Kanto region of Japan. In the present model, which assumes homogeneous parameter values of $\beta$ and $\mu$,  the high-risk area during the daytime is the central business district, where many workers are located. In the nighttime, certain suburban areas from where people commute to the centre become high risk. 
Furthermore, people travelling to the central business district from their suburban residences  allow the disease to spread over the whole metropolitan area.
 Figure \ref{fig:risk}c shows that the infection probability of the test agents can be estimated from the integration of $\rho(\tau,\bm x_p(\tau))$, obtained from stochastic simulation, along its trajectory.

Instead of 
counting the infection probability $r_j(\infty)$ in the neighbourhood of the region of interest, 
 the spectral properties of the averaged adjacency matrix can be incorporated into the risk factor 
by choosing the first few high-contribution modes. 
Namely, one can approximate $r_j(\infty)$ and $\rho(\tau,\bm x)$ as $\tilde r_j^{(M)} (\infty) = \sum_{a=1}^M \hat r_{\sigma(a)} (\infty) \phi_j^{(\sigma(a))} $
and $\tilde \rho^{(M)} (\tau,\bm x) = \sum_{j=1}^N \tilde r_j^{(M)} (\infty) K(\bm x- \bm x_j(\tau))$ using $M$ modes.
Hazard maps based on the mode truncation at different times are similar to those given by the stochastic simulation (Fig.~\ref{fig:risk}a).
Moreover, Fig. \ref{fig:risk}d shows that the integration of $\tilde \rho^{(100)}(\tau,\bm x_p(\tau))$ along the path of the test agents predicts their infection probability.
Thus, we conclude that mode truncation 
enables us to describe the spatio-temporal infection risk with a small number of modes.
In other words, few modes that have high contributions dominate the spread of the epidemic.

\section*{Conclusion}

Data on human mobility, particularly contact information  of urban residents, play a  key role in the analysis of an epidemic spreading.
Such high-resolution data allow detailed  individual-level modelling, rather than coarse-grained metapopulation studies.
At the same time, novel theoretical tools  can be used  to uncover the dominant dynamics 
at work in data-driven models without the need for oversimplification.
Linear analysis 
characterises the exponential growth in the initial stage; nonetheless, this study clarified that the final size of the epidemic is provided by the analysis of various modes.
The  mode truncation described here revealed that the prevalence of epidemics is actually dominated by a small number of modes as compared to the number of agents, i.e. the effective number of degrees of freedom is much smaller than the system size. 
The proposed method practically limits the expanse of mitigation policies such as targeted vaccination, and provides specific alerts by following the spreading trail of infection.
This also quantifies the spatio-temporal risk associated with empirical travel patterns.
The analytical framework and realistic mobility data presented here provide a promising starting point for the study and possible confrontation of epidemics in the real world.
Future studies should examine ways to facilitate the rapid decay of dominant modes using dynamic intervention strategies.

\section*{Methods}

\section*{People flow data and construction of the proximity network}
People flow data were generated based on person--trip surveys made by Japan's Ministry of Land, Infrastructure, Transport, and Tourism. The surveys were based on  questionnaires that asked about basic individual information such as gender, age, occupation, places visited, and trip modes on one day.
For the sake of privacy,  spatial information was recorded with a  resolution such that the complete location could not be determined.
The questionnaires allowed us to access advanced information such as transportation modes, travel times, transfer points, and trip purposes.
Traces of the trips between  two places reported in the raw data were interpolated
using geographical information systems methods 
 \cite{ssku11}.

\begin{figure*}[htbp]
\begin{center}
\begin{minipage}{0.95\hsize}
\includegraphics[width=0.9\textwidth]{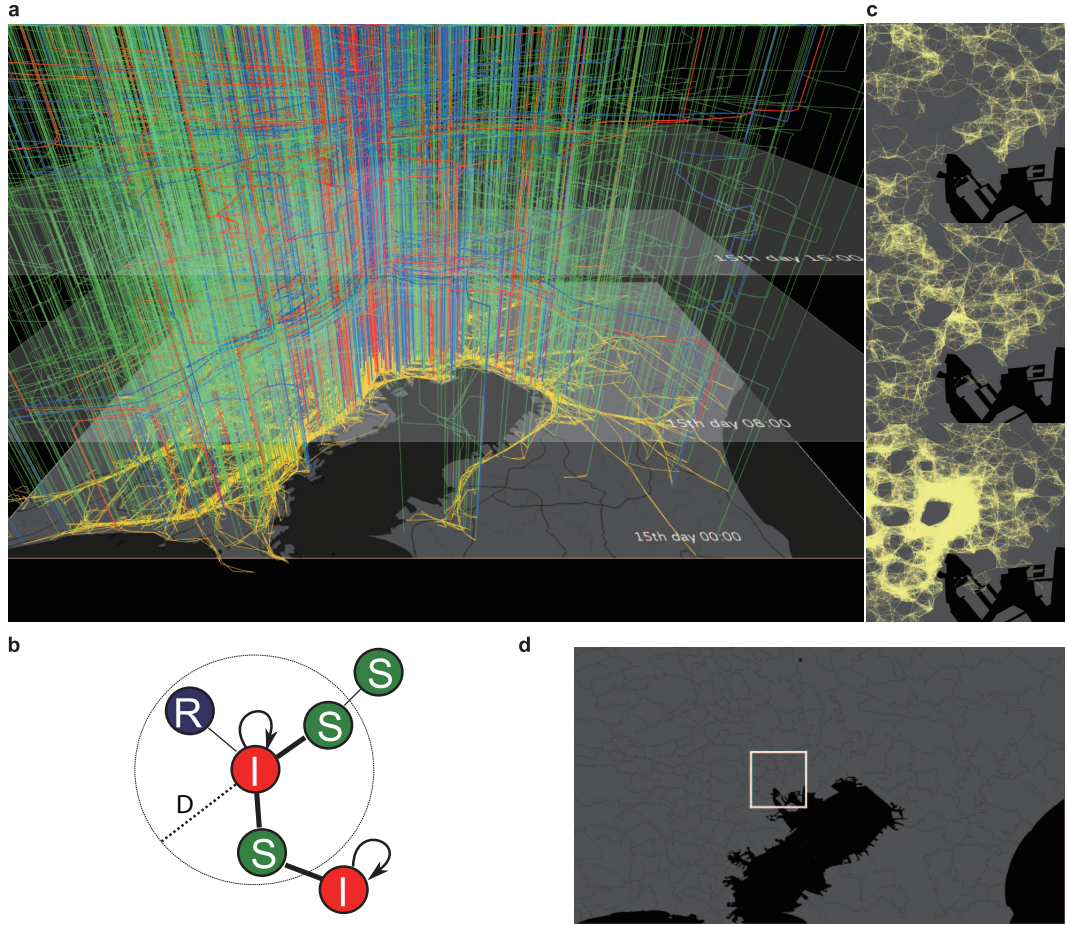}
\end{minipage}
\caption{\label{fig:map}
\textbf{Illustration of epidemic spreading in cities and the time-dependent proximity network.}
a. Mobility pattern and infection state of agents in the Tokyo metropolitan area. Colours represent  susceptible (green), infected (red), and recovered (blue) states on the 15th day of an agent-based stochastic simulation run. The vertical axis represents time.
On the background map, dark grey and black regions represent land and bay areas, respectively. 
Black lines in the land are railways and highways. Yellow lines are the projection of the trajectories.
Density of the yellow lines indicates the traffic density in the region. The centre of the map, which has the highest yellow line density, corresponds to the central business district (CBD) of Tokyo.
As seen from the figure, most infected and recovered agents move to the CBD in the morning and return to their hometowns in the evening.
b. Schematic figure of construction of the proximity network from the People flow data. Infection takes place between susceptible (S) and infected (I) agents within a distance $D$ with  probability $\beta \Delta t$ for a time interval $\Delta t$. Recovered (R) agents do not affect infection of other agents.
c. The connectivity pattern between agents near the CBD for $D=1,000$~m and $N=10,000$ at 00:00, 08:00, and 12:00.
The region of the map is indicated by the white square in the wide-area map in panel d.
}
\end{center}
\end{figure*}

The proximity network was constructed by placing a link between two individuals at time $t$ if they were within the threshold distance $D$ of one another (Fig.~\ref{fig:map}b). 
We added another constraint by only placing a link between individuals using the same transportation mode;  it is impractical to connect individuals walking along the road with those traveling in an automobile or train because of the improbability of the transmission of infection. The connectivity of the network can be described by the time-dependent adjacency matrix $A(t)$, where $A_{jk}(t) = A_{kj}(t) = 1$ if individuals $j$ and $k$ are connected at time $t$ and $A_{jk}(t)=A_{kj}(t) = 0$ otherwise. This allowed  us to obtain networks for $N$  agents at every $\Delta t$ time interval. 
Proximity networks are  spatially embedded temporal networks, 
which appear in various forms of transportation and infrastructure networks \cite{b11}. 

\section*{Settings of the stochastic simulations}
The agent-based SIR model simulation was performed with the same time-dependent adjacency matrix $A(t)$. The time evolution of the probabilities $s_j(t)$, $i_j(t)$, and $r_j(t)$ was computed using 500 independent stochastic simulation runs with different random initial conditions. We also assumed that the human mobility patterns had a periodicity of 24 h owing to the circadian nature and economic situation of people going to their place of work and back home. 
Parameter values were taken as $\Delta t = 10\ \rm{min}$,  $\mu=2.0\times 10^{-4} \ \rm{min}^{-1}$, and $\beta=1.5\times 10^{-5} \ \rm{min}^{-1}$, except for Fig.~\ref{fig:series}d, in which $\beta$ varies.

\section*{Derivation of  equation (\ref{eq:self-conf_eig})}
One of the main results of this article, the equation for the asymptotic solution of equation~(\ref{eq:self-conf_eig}), is derived as follows.
Substituting the expansion (\ref{eq:expansion}) into 
equation~(\ref{eq:mf}), we obtain
\begin{align}
\frac{d s_j}{dt}&= - \beta s_j \sum_{a=1}^N \lambda_a \hat i_a(t) \phi _j^{(a)}.
\end{align}
Let us define
\begin{align}
\Psi_j(t) 
& \equiv \sum_{a=1}^N \int_0^t  \lambda_a \hat i_a(t') \phi_j^{(a)} dt' \\
&= \frac{1}{\mu} \sum_{a=1}^N \lambda_a   [\hat r_a(t) - \hat r_a(0)] \phi_j^{(a)}, \label{eq:Psi-r}
\end{align}
so that the solution of this equation is given as
\begin{align}
s_j (t) &= s_j(0) e^{-\beta \Psi_j(t)} ,
\end{align}
or equivalently, 
\begin{align}
\hat s_a (t) &= \sum_{a=1}^N s_j(0) \phi_j^{(a)}  e^{-\beta \Psi_j(t)}. \label{eq:s_evo}
\end{align}

For $t\rightarrow \infty$, $i_j(\infty)$ should vanish,
and the conservation of probability condition is given as
$s_j(\infty) + r_j(\infty) = 1$.
This is equivalent to 
\begin{align}
\hat r_a(\infty) &=  N\langle \phi^{(a)} \rangle - \hat s_a(\infty) .
\end{align}
Substituting equation~(\ref{eq:s_evo}) into this expression, 
we obtain 
 \begin{align}
\hat r_{a}(\infty) 
=
N\langle  \phi^{(a)} \rangle  -  \sum_{j=1}^N s_j(0) \phi_j^{(a)} 
\exp \left\{ -\frac{\beta}{\mu}  \sum_{b=1}^N \lambda_{b} [\hat r_{b} (\infty) - \hat r_{b}(0)] \phi_j^{(b)} \right\}.
    \label{eq:self-conf_eig_orig} 
\end{align}
This is a transcendental equation for $\hat r_a(\infty)$,
and one can determine the infection probability by solving this equation
instead of solving the differential equation.
We choose $M$ ($\le N$) modes and neglect the other modes to obtain
 \begin{align}
\hat r_{\sigma(a)}(\infty) 
\approx 
  N\langle  \phi^{(\sigma(a))} \rangle  -  \sum_{j=1}^N s_j(0) \phi_j^{(\sigma(a))} 
\exp \left\{-\frac{\beta}{\mu}  \sum_{b=1}^M \lambda_{\sigma(b)} [\hat r_{\sigma(b)} (\infty) - \hat r_{\sigma(b)}(0)] \phi_j^{(\sigma(b))} \right\},
    \label{eq:self-conf_eig_orig2} 
\end{align}
where $\sigma(a)$ and $\sigma(b)$ denote arbitrary permutations of modes.
Finally, we multiply both sides by $\langle \phi^{(\sigma(a))} \rangle $, and obtain equation~(\ref{eq:self-conf_eig}) by choosing the top $M$ modes with respect  to their contribution.

\section*{Derivation of the final size equation in heterogeneous mean field theory}
The relationship between the equation for the expansion coefficients (\ref{eq:self-conf_eig_orig}) and that 
obtained with HMF can be described as follows.
We concentrate on the simplest case, where the probability $P_{dd'}$ that
a node with degree $d$ and one with degree $d'$ are connected is
proportional to the product of their degrees, i.e. $P_{dd'} = d d' P(d') / \langle d \rangle $, where $P(d)$ denotes the degree distribution of the network satisfying $\sum_{d=1}^{N-1} P(d) = 1$.
In this case, the largest eigenvalue of the adjacency matrix is
$\lambda_1 = \langle d^2 \rangle / \langle d \rangle $,
and the component of the corresponding normalised eigenvector is $\phi_j^{(1)} =d_j/\sqrt{N \langle d^2 \rangle }$.
Here, we denote the degree of node $j$ as $d_j$.
We assume that the probability of nodes with the same degree is identical, i.e.
$r_j(t) = \overline{r}_{d_j}(t)$ holds.

Substituting the expression for the first eigenvector into
equation~(\ref{eq:self-conf_eig_orig}) and neglecting other modes with the initial condition 
$s_j(0) \approx 1$, we get
\begin{align}
\hat r_1(\infty) 
& \approx
  N\langle  \phi^{(1)} \rangle  -  \sum_{j=1}^N  \phi_j^{(1)} e^{-\frac{\beta}{\mu}  \lambda_1 \hat r_1 (\infty)  \phi_j^{(1)}}  \\
  &=
\sqrt{\frac{N}{ \langle d^2 \rangle }}
\left[ \langle  d \rangle - \sum_{d=1}^{N-1} d P(d)  e^{-\frac{\beta}{\mu}  \langle d^2 \rangle  \hat r_1 (\infty)  d / (\langle d \rangle \sqrt{N \langle d^2 \rangle} )}
\right] ,
\label{eq:final_dd}
\end{align}  
where the sum in the second equation is taken for degree $d$.
Note that 
\begin{align}
\hat r_1(t) &= \sum_{j=1}^N r_j(t) \phi_j^{(1)}= \sqrt{\frac{N}{\langle d^2 \rangle }}\sum_{d=1}^{N-1} dP(d) \overline{r}_d(t) \\
 &= \sqrt{\frac{N}{\langle d^2 \rangle }} \mu \langle d \rangle \Psi(t)
\end{align}
holds, where we have defined 
\begin{align}
\Psi(t) 
& \equiv
\frac{1}{\mu \langle d \rangle} \sum_{d=1}^{N-1} d P(d) \overline{r}_{d} (t).
\end{align}
Substituting this equation into equation~(\ref{eq:final_dd}) and taking the limit $t\rightarrow \infty$,
we obtain
\begin{align}
\mu \Psi (\infty)  &= 1 
- \frac{1}{ \langle d \rangle} \sum_{d=1}^{N-1} d P(d) e^{-\beta d \Psi(\infty)} .
\end{align}
This is the equation for $\Psi(\infty)$ in HMF derived in \cite{mpv02}.

\section*{Acknowledgement}
N.F. is grateful for the stimulating discussions with T. Takaguchi, T. Aoki, Y. Sughiyama, and Y. Yazaki.
This research is supported by the Aihara Project, the FIRST program from JSPS, initiated by CSTP, and  by CREST, JST.
This research is the result of the joint research with Center for Spatial Information Science, the University of Tokyo (No. 315). 
N.F. is supported by JSPS KAKENHI Grant Number 15K16061.

\section*{Author contributions}
All authors designed the research. N.F. and A.R.S. carried out the numerical simulations. N.F., A.R.S, and K.I. analysed the results. N.F., K.I., and K.A. contributed to the analytical calculations.
N.F. and A.R.S. wrote the bulk of the manuscript. K.I. and K.A. edited the manuscript.

\section*{Competing financial interests}
The authors declare no competing financial interests.

\clearpage
\thispagestyle{empty}

\clearpage
\thispagestyle{empty}

\clearpage
\thispagestyle{empty}

\clearpage
\thispagestyle{empty}

\thispagestyle{empty}

\clearpage

\begin{center}
\section*{\large Supplementary information: Contribution of hidden modes to nonlinear epidemic dynamics in urban human proximity networks}
\end{center}

\section{Scaling of the giant cluster size}
\begin{figure}[htbp]
\begin{center}
\begin{minipage}{0.45\hsize}
\includegraphics[width=0.95\textwidth]{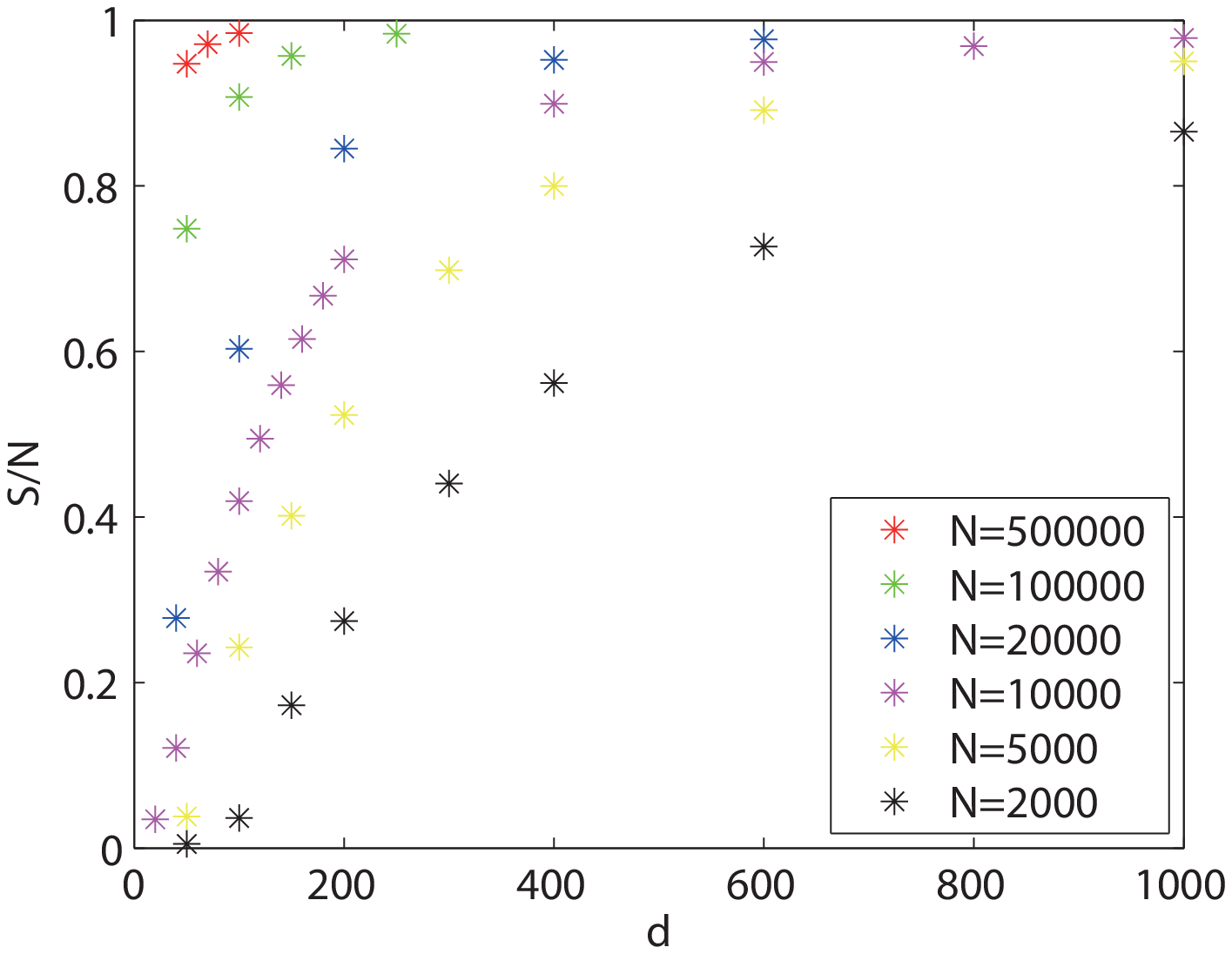}
\end{minipage}
\begin{minipage}{0.45\hsize}
\includegraphics[width=0.95\textwidth]{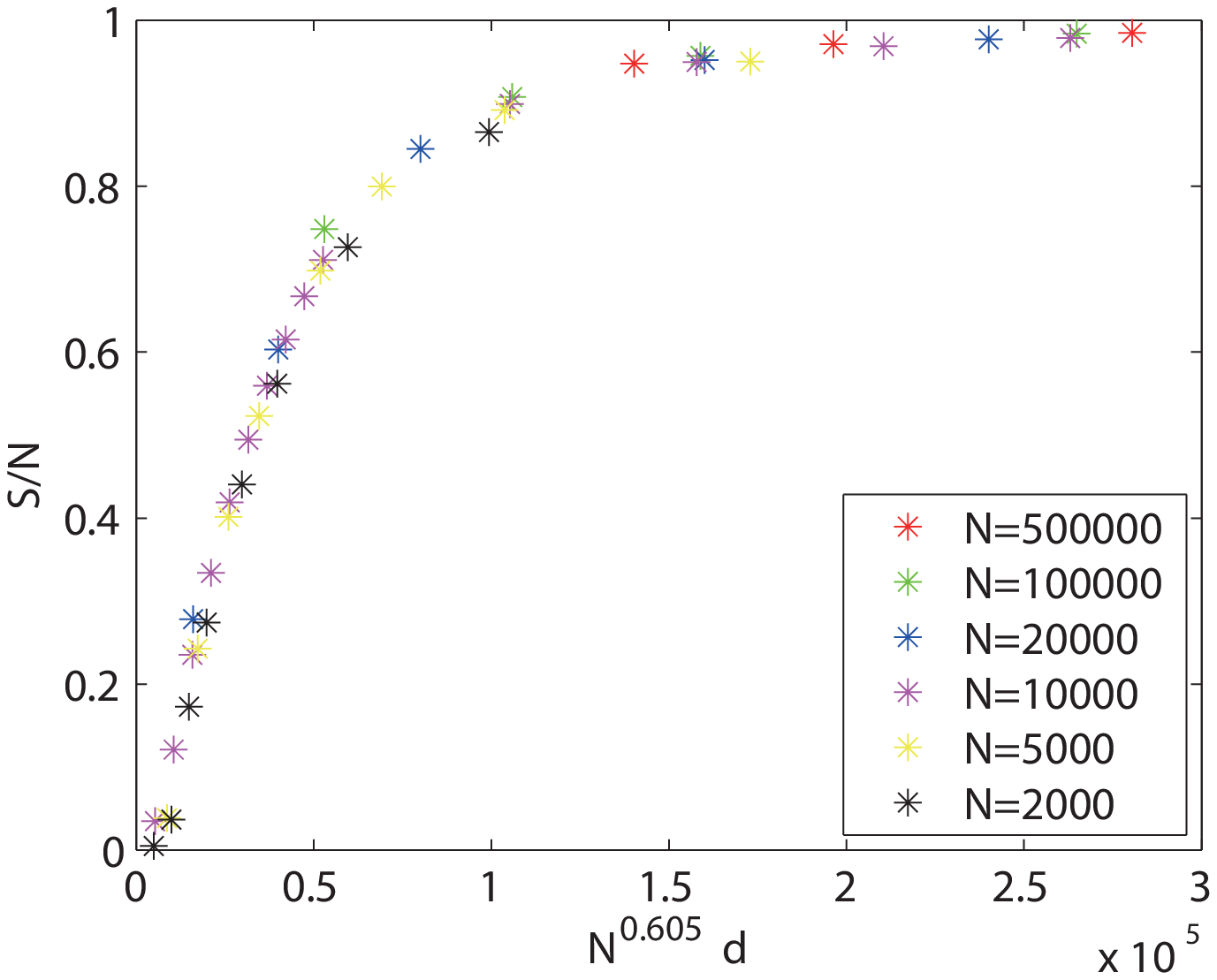}
\end{minipage}
\caption{
\label{fig:GCC}
Scaling of the giant cluster component size ratio for different $N$.
The right-hand figure shows that this is a function of the scaled parameter $N^{0.605} d$.
}
\end{center}
\end{figure}

The data size of the People flow data of $N\approx 500,000$ is much smaller than the population of the corresponding Tokyo metropolitan area, $N\approx 27,500,000$.
Therefore, it is  natural to ask whether it makes sense to consider a model with a small number of agents.
A clue to answering this question is to use scaling.
We found that this system exhibits a percolation-like transition:
the order parameter of the giant cluster component size ratio $S(N,d)/N$ has a second-order phase transition (Fig.~\ref{fig:GCC} left).
Although it is unclear whether the percolation-like transition takes place for large $N$
 (because of the spatial resolution of the data for privacy reasons), 
we found that $S(N,d)/N$ obeys the scaling law (Fig.~\ref{fig:GCC} right).
Far from the percolation transition point, $S(N,d)/N \sim f(N^{0.605} d) $ holds, where $f$ is a scaling function.
If we assume that this scaling holds for the size of the real population, the parameter values $N=10,000$, $d=1,000$~m used in the simulations in the main text correspond to $d \sim10,000^{0.605} \times 1,000 / 27,500,000^{0.605}\approx 8.3$~m in the real population.

\section{Time-averaged networks}
We discuss an approximation to replace the time-dependent adjacency matrix with the time-averaged one, and determine the condition under which this approximation is valid. Let us consider  a general example where a dynamical variable $\bm X(t)$
 evolves via contact with other nodes 
 as
\begin{align}
\dot{\bm X} (t)&=  \beta A(t)  \bm X(t) ,
\label{eq:matrix_ode}
\end{align}
where the parameter $\beta$ represents the coupling strength and $A(t)$ 
is  the instantaneous adjacency matrix 
of a temporal network 
at time $t$.
Equation (\ref{eq:matrix_ode}) can be solved as
\begin{align}
\bm X(t) &= \exp_{\rm T} \left[ \beta \int_0^t dt' A(t') \right] \bm X(0),
\label{eq:tprod}
\end{align}
where $\exp_{\rm T}$ denotes the time-ordered exponential. 
If $A(t)$ changes with the time interval $\tau$, 
equation (\ref{eq:tprod}) can be rewritten  as
\begin{align}
\bm X (L\tau) &=  e^{\beta \tau A(t+(L-1)\tau)} e^{\beta \tau A(t+(L-2) \tau)} \cdots e^{\beta \tau A(\tau) } e^{\beta \tau A(0)} \bm X(0) \\
&= (1+\beta L \tau  \overline{A} ) \bm X(0) +O(\beta^2),
\label{eq:fsa}
\end{align}
where 
$\overline{A} =   \sum_{l=0}^{L-1} A(t+l\tau)/L$ is  the time-averaged adjacency matrix.
The  above equation can be interpreted as follows.
The terms of  $\beta^m$ in the power series expansion in equation (\ref{eq:tprod}) consist of  time-ordered products of $m$ adjacency matrices of the temporal network.
These terms collect paths of length $m$,  namely  all possible 
indirect interactions occurring $m$ times between two nodes
at different times taking into account the time order of the interactions.
If we consider the case of a small value of  $\beta$, 
the time evolution can be approximated up to the first order. 
This corresponds to assuming that the interaction only takes place 
 by direct contact between agents, and all multi-step interactions during the time interval $T=L\tau$ are negligible.
In such a case, the time-averaged network is 
sufficient to describe the dynamical processes in the temporal network.
Note that the time-order information from one day does not appear in the first-order terms.

\section{Nonlinear mode truncated  equations}
One can derive the mode coupling equation from the following differential equations (1)  in the main text: 
\begin{align}
\begin{split}
\frac{d s_j}{dt} &= - \beta s_j \sum_{k=1}^N \overline{A}_{jk} i_k \\
\frac {d i_j}{dt} &= \beta s_j \sum_{k=1}^N \overline{A}_{jk} i_k -\mu i_j \\
\frac{d  r_j}{dt} &= \mu i_j.
\label{eq:mf}
\end{split}
\end{align}
Let us substitute equation (2) in the main text,
\begin{align}
\begin{split}
s_j(t) = \sum_{a=1}^N \hat s_a(t) \phi^{(a)}_j,\ 
i_j (t) = \sum_{a=1}^N \hat i_a(t) \phi^{(a)}_j,\ 
r_j (t) = \sum_{a=1}^N \hat r_a (t) \phi^{(a)}_j,
\end{split}
\label{eq:expansion}
\end{align}
 into this equation.
Taking into account the eigenvalue equation for the eigenvector
 $\phi_j^{(a)}$,
\begin{align}
\sum_{k=1}^N A_{jk} \phi^{(a)}_k = \lambda_a \phi^{(a)}_j,
\end{align}
it is easy to verify that
\begin{align}
\begin{split}
\frac{d }{dt}  \bigg[\sum_{a'=1}^N \hat s_{a'} \phi^{(a')}_j \bigg]
&= 
- \beta \bigg( \sum_{b=1}^N \hat s_b \phi^{(b)}_j \bigg) \bigg[ \sum_{k=1}^N  
\overline{A}_{jk}(t) \bigg( \sum_{c=1}^N \hat i_c \phi^{(c)}_k \bigg) \bigg]\\
&= 
- \beta \bigg( \sum_{b=1}^N \sum_{c=1}^N \lambda_c 
   \phi^{(b)}_j    \phi^{(c)}_j  \hat s_b  \hat i_c \bigg),\\
\frac {d }{dt}
 \bigg[\sum_{a'=1}^N \hat i_{a'} \phi^{(a')}_j \bigg]
 &= 
 \beta \bigg( \sum_{b=1}^N \hat s_b \phi^{(b)}_j \bigg) \bigg[ \sum_{k=1}^N  
\overline{A}_{jk}(t) \bigg( \sum_{c=1}^N \hat i_c \phi^{(c)}_k \bigg) \bigg]
 -\mu \sum_{a'=1}^N \hat i_{a'} \phi_j^{(a')} \\
 &=
 \beta \bigg( \sum_{b=1}^N \sum_{c=1}^N \lambda_c 
   \phi^{(b)}_j    \phi^{(c)}_j  \hat s_b  \hat i_c \bigg)
 -\mu \sum_{a'=1}^N \hat i_{a'} \phi_j^{(a')} ,\\
\frac{d  }{dt}
 \bigg[\sum_{a'=1}^N \hat r_{a'} \phi^{(a')}_j \bigg]
 &=  \mu \sum_{a'=1}^N \hat i_{a'} \phi_j^{(a')} .
\label{eq:mf2}
\end{split}
\end{align}
Multiplying $\phi_j^{(a)}$ on both sides of the above equations and taking the sum over $j$
to use the orthonormal condition
\begin{align}
\sum_{j=1}^N \phi_j^{(a)} \phi_j^{(b)} = \delta_{ab},
\end{align}
where $\delta_{ac}$ is Kronecker's delta,
 we obtain 
\begin{align}
\begin{split}
\frac{d \hat s_a}{dt} 
&= 
- \beta \bigg( \sum_{b=1}^N \sum_{c=1}^N  \lambda_c   G_{abc}  \hat s_b  \hat i_c \bigg),\\   
\frac {d \hat i_a}{dt}
 &= 
\beta \bigg( \sum_{b=1}^N \sum_{c=1}^N  \lambda_c   G_{abc}  \hat s_b  \hat i_c \bigg)
 -\mu \hat i_a ,\\
\frac{d  \hat r_a}{dt}
 &= \mu  \hat i_a  ,
\label{eq:mf3}
\end{split}
\end{align}
where $\displaystyle G_{abc} = \sum_{j=1}^N \phi_j^{(a)} \phi_j^{(b)} \phi_j^{(c)}$. Note that each element of $G_{abc}$ consists of the product of the eigenvector components and is time-independent.

It is easy to extend this equation to the case where the static adjacency matrix $\overline{A}$ is asymmetric.
Let us define the left and right eigenvectors of $\overline{A}$ for the eigenvalue $\lambda_a$ as
$\psi_j^{(a)}$ and $\phi_j^{(a)}$, and set the normalisation condition $\sum_{j=1}^N \psi_j^{(a)} \phi_j^{(b)} = \delta_{ab}$.
Then, equation (\ref{eq:mf3}) holds if we define
$ \displaystyle  G_{abc} = \sum_{j=1}^N \psi_j^{(a)} \phi_j^{(b)} \phi_j^{(c)}$. 

By considering $M$ ($\le N$) modes and neglecting the other modes in equation (\ref{eq:mf3}), we obtain the transient dynamics of the epidemic from a small number of modes.

\section{Derivation of the linearised equation from the mode coupling equation}
Equations (\ref{eq:mf3}) contain mode coupling terms, but the modes are decoupled in the linearised equation.
Let us take $s_j(t) = 1 - \Delta s_j(t) $ and focus on the initial stage of the epidemic spreading. 
Assume that $\Delta s_j(t)$ and $i_j(t)$ are small, and approximate $s_j i_j \approx i_j$ by neglecting the product $\Delta s_j i_j$. 
Since $s_j$ appears only as the product $s_j i_j$, the linearised approximation allows us to put $s_j \approx 1$.
Under this assumption, we can take 
\begin{align}
 s_j = \sum_{b=1}^N \hat s_b \phi_j^{(b)} = 1.
 \label{eq:approx_mode}
 \end{align}
Substituting this into the evolution equation for $\hat i_a$ in equation (\ref{eq:mf3}), the linearised equation becomes
\begin{align}
\frac {d \hat i_a}{dt}
 & \approx
\beta \bigg( \sum_{b=1}^N \sum_{c=1}^N  \lambda_c   G_{abc}    \hat s_b \hat i_c \bigg)
 -\mu \hat i_a \\
 & =
\beta \bigg( \sum_{j=1}^N \sum_{b=1}^N \sum_{c=1}^N  \lambda_c   \phi_j^{(a)} \phi_j^{(b)} \phi_j^{(c)}    \hat s_b \hat i_c \bigg)
 -\mu \hat i_a .
\end{align}
By taking the summation over $b$ using equation (\ref{eq:approx_mode}), we obtain
\begin{align}
\frac {d \hat i_a}{dt}
 & \approx
\beta \bigg(\sum_{j=1}^N \sum_{c=1}^N  \lambda_c   \phi_j^{(a)} \phi_j^{(c)}     \hat i_c \bigg)
 -\mu \hat i_a \\
 &=
\beta \bigg( \sum_{c=1}^N  \lambda_c      \hat i_c   \delta_{ac} \bigg)
 -\mu \hat i_a \\ 
 &=
( \beta \lambda_a - \mu) \hat i_a.
\end{align}
In this way, we derive equation  (3) in the main text.

\section{Identification of hot spots and estimation of the infection probability of a test agent}
We now discuss how to estimate the infection probability of a test agent.
Let $\bm x_j (t) = (x_j(t), y_j(t) )^T $ be the position of agent $j$ at time $t$.
Since the infection probability depends on the distance between agents, we introduce 
the kernel function $K(\bm x)$.
In this study, we used the Heaviside step function $K(\bm x) = \Theta ( |\bm x| -D)$, where $\Theta(x) = 1$ for $x \ge 0$ and $\Theta (x) = 0$ otherwise.
Then the element of the adjacency matrix $A_{jk}(t)$ is given as $A_{jk} (t) = K(\bm x_j(t)  - \bm x_k(t))$.

We define the spatio-temporal risk factor of infection as
\begin{align}
\rho (\tau, \bm x) \equiv   \sum_{k=1}^N r_k (\infty) K\big(\bm x - \bm x_k(\tau) \big).
\end{align}
where we take $0 \le \tau \le T = 1440 \ \rm{min}$, because
we assume the mobility pattern is periodic over 24 h and $\rho(\tau +T,\bm x)=\rho (\tau, \bm x)$ holds.

We want to estimate the risk of infection of a test agent
$p$ who travels inside the urban area with an arbitrary trajectory and returns to the original location after time $T$. 
{Equation (1) in the main text}
can be solved as
\begin{align}
s_j (t) &= s_j(0) e^{-\beta \Psi_j(t)} ,
\end{align}
where
\begin{align}
\Psi_j(t) 
&=  \int_0^t  \sum_{k=1}^N \overline A_{jk} i_k(t') dt' \\
&= \frac{1}{\mu} \sum_{k=1}^N \overline{A}_{jk}  [ r_k(t) - r_k(0)]. 
\end{align}
For $t\rightarrow \infty$, $i_j(\infty)$ should vanish,
and the conservation of probability condition is given as
$ r_j(\infty) = 1 - s_j(\infty) $.
Therefore, the infection probability of this agent is 
\begin{align}
r_j(\infty) &= 1-s_j(0) \exp \left\{ -\frac{\beta}{\mu} \sum_{k=1}^N \overline{A}_{jk} [r_k (\infty) - r_k(0)] \right\}.  \label{eq:final}
\end{align}

This equation is  transcendental and  cannot  be solved analytically. 
However, if $N$ is sufficiently large  and the contribution of the test agent to $\rho(\tau,\bm x)$ is negligible,
we can approximate the infection probability  of the test agent $p$ using $\rho(\tau,\bm x)$.
Since the time-averaged adjacency matrix is given in terms of the risk factor as
\begin{align}
\sum_{k=1}^N \overline{A}_{pk} r_k(\infty) &=  \displaystyle \frac{1}{T} \sum_{k=1}^N \int_0^T  A_{pk}(\tau) r_k(\infty) d\tau \\
&= 
\frac{1}{T} \int_0^T \rho(\tau,\bm x_p(\tau)) d\tau,
\end{align}
we obtain
\begin{align}
r_p(\infty) &= 1- \exp \left[-  \frac{\beta}{\mu T}  \int_0^T \rho(\tau,\bm x_p(\tau) ) d\tau \right],
\label{eq:final_add}
\end{align}
by taking $s_j(0) =1$ and $ r_k(0) = 0$.
Therefore, the final infection probability of a susceptible agent traveling along an arbitrary trajectory is given by the time integral of the risk factor along its trajectory.

If the mode truncation with $M$ ($\le N$) modes
\begin{align}
r_j (t) 
&\approx \tilde r_j^{(M)} (t) \equiv \sum_{a=1}^{M} \hat r_{\sigma(a)} (t) \phi_j^{(\sigma(a))}
\end{align}
approximates $r_j(t)$,
the risk factor 
is  given by the superposition of $M$ independent variables as
\begin{align}
\rho(\tau,\bm x) &\approx\tilde \rho^{(M)} (\tau,\bm x) \equiv  \sum_{a=1}^M \sum_{j=1}^N \hat r_{\sigma(a)} (\infty) \phi_j^{(\sigma(a))} K(\bm x-\bm x_j(\tau)).
\end{align}
The infection probability  is approximated with the mode truncation 
\begin{align}
r_p(\infty) &\approx \tilde r_p(\infty) = 1- \exp \left[ -
\frac{\beta}{\mu T} \int_0^T \tilde \rho^{(M)}(\tau,\bm x_p(\tau)) d\tau \right] ,
\end{align}
by replacing $\rho (\tau,\bm x)$ with $\tilde \rho^{(M)}(\tau,\bm x)$ in equation (\ref{eq:final_add}).

\end{document}